\documentclass[a4paper, amsfonts, amssymb, amsmath, reprint, showkeys, nofootinbib, twoside, floatfix]{revtex4-1}
\usepackage[english]{babel}
\usepackage[utf8]{inputenc}
\usepackage[colorinlistoftodos, color=green!40, prependcaption]{todonotes}
\usepackage{amsthm}
\usepackage{mathtools}
\usepackage{physics}
\usepackage{xcolor}
\usepackage{graphicx}
\usepackage[left=23mm,right=13mm,top=35mm,columnsep=15pt]{geometry} 
\usepackage{adjustbox}
\usepackage{placeins}
\usepackage[T1]{fontenc}
\usepackage{lipsum}
\usepackage{csquotes}
\usepackage{subfiles}
\usepackage[pdftitle={Article}, pdfauthor={Author}]{hyperref} % For hyperlinks in the PDF
\usepackage{graphicx}
\usepackage[caption=false]{subfig}
\usepackage[left]{lineno}

\begin{document}

\title{Direct limits for scalar field dark matter from a %quantum-enhanced 
gravitational-wave detector}

\author{Sander M. Vermeulen$^1$}
\author{Philip Relton$^1$}
\author{Hartmut Grote$^1$}
\email[Correspondence email address: ]{groteh@cardiff.ac.uk}
\author{Vivien Raymond$^1$}
\author{Christoph Affeldt$^2$}
\author{Fabio Bergamin$^2$}
\author{Aparna Bisht$^2$}
\author{Marc Brinkmann$^2$}
\author{Karsten Danzmann$^2$}
\author{Suresh Doravari$^2$}
\author{Volker Kringel$^2$}
\author{James Lough$^2$}
\author{Harald L\"uck$^2$, Moritz Mehmet$^2$, Nikhil Mukund$^2$, S\'everin Nadji$^2$, Emil Schreiber$^2$, Borja Sorazu$^3$, Kenneth A. Strain$^{2,3}$, Henning Vahlbruch$^2$}
\author{Michael Weinert$^2$}
\author{Benno Willke$^2$}
\author{Holger Wittel$^2$}
\affiliation{$^1$Gravity Exploration Institute, Cardiff University, Cardiff CF24 3AA, United Kingdom}
\affiliation{$^2$Max-Planck-Institute for Gravitational Physics and Leibniz University Hannover, Callinstr. 38, 30167 Hannover, Germany}
\affiliation{$^3$School of Physics and Astronomy, University of Glasgow, Glasgow G12 8QQ, United Kingdom}
\date{\today} % Leave empty to omit a date

\begin{abstract}
The nature of dark matter remains unknown to date; several candidate particles are being considered in a dynamically changing research landscape~\cite{Bertone_2018}.
Scalar field dark matter is a prominent option that is being explored with precision instruments, such as atomic clocks and optical cavities~\cite{arvanitaki_searching_2015,derevianko_hunting_2014,stadnik_can_2015,van_tilburg_search_2015,hees_searching_2016,leefer_search_2016,savalle_searching_2021}.
Here we report on the first direct search for scalar field dark matter utilising a gravitational-wave detector,
which operates beyond the quantum shot-noise limit.
We set new upper limits for the coupling constants of scalar field dark matter as a function of its mass, by excluding the presence of signals that would be produced through the direct coupling of this dark matter to the beamsplitter of the GEO\,600 interferometer. The new constraints improve upon bounds from previous direct searches by more than six orders of magnitude, and are in some cases more stringent than limits obtained in tests of the equivalence principle by up to four orders of magnitude.
Our work demonstrates that scalar field dark matter can be probed or constrained with direct searches using gravitational-wave detectors, and highlights the potential of quantum-enhanced interferometry for dark matter detection.

%The nature of dark matter remains unknown to date and several candidate particles are being considered.
%In this work we report on the first direct search for scalar field dark matter in a quantum-enhanced gravitational-wave detector. We set new upper limits for the coupling constants of scalar field dark matter as a function of its mass by excluding the presence of signals produced through the direct coupling of this candidate dark matter to the beamsplitter of the GEO\,600 interferometer. The new constraints improve upon bounds from previous direct searches by more than six orders of magnitude, and are more stringent than limits obtained in tests of the equivalence principle by one order of magnitude.  
%%We set new upper limits for scalar field dark matter by exploring their direct coupling to the GEO\,600 interferometer.
\end{abstract}

%\keywords{dark matter, gravitational wave detector}

\maketitle

\section{Introduction} \label{sec:introduction}
Laser interferometers have exquisite sensitivity to minute length changes of space, and have facilitated many gravitational-wave detections over the last years~\cite{GWTC1_2019,GWTC2_2020}.
In addition to their revolutionary merit in astrophysics, the detection of gravitational waves has also shed light on fundamental physics questions, and several links may exist between gravitational waves and dark matter~\cite{Bertone2019}. 
Due to their excellent sensitivity at or beyond quantum limits, gravitational-wave detectors (or precision interferometers of a similar type) can be used directly for fundamental physics, without the mediation of gravitational waves. Examples include a possible search for vacuum birefringence~\cite{Grote2015}, and the search for signatures of quantum gravity~\cite{Chou2016,Verlinde2019,vermeulen_experiment_2021}. Several ideas have been put forward as to how different candidates of dark matter can directly couple to gravitational-wave detectors, ranging from scalar field dark matter~\cite{stadnik_can_2015,Grote2019} to dark photon dark matter~\cite{pierce_searching_2018}, and to clumpy dark matter coupling gravitationally or through an additional Yukawa force~\cite{Hall2018}. Upper limits for dark photon dark matter have already been set in a small mass band using data from the first observational run (O1) of the Advanced LIGO gravitational-wave detectors~\cite{Zhao2019}.

%In this work we demonstrate how a quantum-enhanced gravitational-wave detector~\cite{GEO6dB_2021} is used to set new upper limits on scalar field dark matter, which is the first direct search for dark matter of this kind with a gravitational-wave interferometer.
In this work we conduct the first direct search for scalar field dark matter using a gravitational-wave detector, the quantum-enhanced GEO\,600 interferometer, and set new upper limits on the parameters of such dark matter.
% this work we set new upper limits on scalar field dark matter using a quantum-enhanced gravitational-wave detector~\cite{GEO6dB_2021}, which is the first direct search for dark matter of this kind with a gravitational-wave interferometer. 
  %I believe leaving the sections in separate files is more organized, change it if you desire 
\section{Theory}\label{sec:theory}
% \begin{equation}
%     \phi(t,\vec{r}) = \phi_0 \cos\left(f_\phi t - \vec{k}_\phi \cdot \vec{r}\right),
% \end{equation}

Models of weakly coupled low-mass ($\ll 1\, \mathrm{eV}$) scalar fields predict that such particles could be produced in the early Universe through a vacuum misalignment mechanism, and would manifest as a coherently oscillating field \cite{arvanitaki_searching_2015,stadnik_can_2015}, 
\begin{equation}\label{eq:osc_field}
        \phi(t,\vec{r}) = \phi_0 \cos\left(\omega_\phi t - \vec{k}_\phi \cdot \vec{r}\right),
\end{equation}
where $\omega_\phi = %\frac{m_\phi c^2}{2\pi\hbar}
(m_\phi c^2) / \hbar$ is the angular Compton frequency, and $\vec{k}_\phi = %\frac{m_\phi \vec{v}_{\text{obs}}}{\hbar}
(m_\phi\vec{v}_{\text{obs}} )/ \hbar $ is the wave vector, with $m_\phi$ the mass of the field and $\vec{v}_{\text{obs}}$ the velocity relative to the observer. The amplitude of the field can be set as $\phi_0 = % \frac{\sqrt{2\rho_{\mathrm{LDM}}}}{m_\phi}m_\phi
(\hbar \sqrt{2 \rho_{\mathrm{local}}}) / (m_\phi c)$, under the assumption that this scalar field constitutes the local dark matter (DM) density $\rho_{\mathrm{local}}$ \cite{read_local_2014}.  
%Non-zero velocities produce a Doppler-shift, which changes the observed DM field frequency.
% Non-zero velocities produce a Doppler-shift, giving an observed DM field frequency
% \begin{equation}
%      \omega_\text{obs} = \omega_\phi + \frac{m_\phi \vec{v}_{\text{obs}}^2}{2\hbar}.
% \end{equation} 

Moreover, these models predict such DM would be trapped and virialised in gravitational potentials, leading to a Maxwell-Boltzmann-like distribution of velocities $\vec{v}_{\text{obs}}$ relative to an observer. As non-zero velocities produce a Doppler-shift of the observed DM field frequency, this virialisation results in the DM field having a finite coherence time or, equivalently, a spread in observed frequency (linewidth) %\frac{\Delta \omega_{\text{obs}}}{\omega_{\text{obs}}}\approx 10^{-6}
%${\Delta\omega_{\text{obs}}/\omega_{\text{obs}}\sim 10^{-6}}$
\cite{derevianko_detecting_2018,pierce_searching_2018}. The linewidth is determined by the virial velocity, which is given by the depth of the gravitational potential. For DM trapped in the galactic gravity potential, as in the standard galactic DM halo model, the expected linewidth is ${\Delta\omega_{\text{obs}}/\omega_{\text{obs}}\sim 10^{-6}}$. Certain kinds of scalar particles, such as Relaxion DM~\cite{flacke_phenomenology_2017,banerjee_coherent_2019}, may also form gravitationally bound objects and be captured in the gravitational potential of the Earth or Sun, producing a local DM overdensity where the field has a much narrower linewidth~\cite{banerjee_relaxion_2020}. % A relaxion halo centred on Earth for example, would entail a DM field with a linewidth smaller than that of galactic halo DM by 3 orders of magnitude, and a density greater by up to 16 orders of magnitude \cite{?}. 
The observed DM frequency is further modulated by the motion of the Earth with respect to the local DM's centre of mass.     

Scalar field DM could couple to the fields of the Standard Model (SM) in numerous ways. Such a coupling, sometimes called a `portal', is modelled by the addition of a parameterised interaction term to the SM Lagrangian~\cite{ringwald_exploring_2012,hees_violation_2018}. In this paper, we consider linear interaction terms with the electron rest mass $m_e$ and the electromagnetic field tensor $F_{\mu\nu}$:
\begin{equation}\label{L_int}
    \mathcal{L}_\mathrm{int} \supset \frac{\phi}{\Lambda_\gamma} \frac{F_{\mu\nu}F^{\mu\nu}}{4}  - \frac{\phi}{\Lambda_e} m_e \bar{\psi}_e \psi_e,
\end{equation}
where $\psi_e$, $\bar{\psi}_e$ are the SM electron field and its Dirac conjugate, and $\Lambda_\gamma$, $\Lambda_e$ parameterise the coupling. Specific types of scalar DM, such as the hypothetical Moduli and Dilaton fields motivated by string theory, have couplings to the QCD part of the SM as well~\cite{damour_string_1994,arvanitaki_sound_2016,damour_equivalence_2010}. 

The addition of the terms in Eq.~\ref{L_int} to the SM Lagrangian entails changes of the fine structure constant $\alpha$ and the electron rest mass $m_e$~\cite{derevianko_hunting_2014,arvanitaki_searching_2015}.
% \begin{equation}
%     \frac{\delta \alpha}{\alpha}= \frac{\phi}{\Lambda_\gamma}, \qquad \frac{\delta m_e}{m_e}= \frac{\phi}{\Lambda_e},
% \end{equation}
% to first order. 
The apparent variation of these fundamental constants in turn changes the lattice spacing and electronic modes of a solid, driving changes of its size $l$ and refractive index $n$:
\begin{align}\label{delta_l}
    \frac{\delta l}{l}&= - \left( \frac{\delta \alpha}{\alpha} +   \frac{\delta m_e}{m_e}\right), \\ \frac{\delta n}{n}&= - 5\cdot 10^{-3}\left(2 \frac{\delta \alpha}{\alpha} +   \frac{\delta m_e}{m_e}\right), \label{delta_n}
\end{align}
where $\delta x$ denotes a change of the parameter $x$: ${x\rightarrow x + \delta x}$. Eqs. \ref{delta_l}, \ref{delta_n} hold in the adiabatic limit, which applies for a solid with a mechanical resonance frequency much higher than $\omega_\phi$ (the driving frequency)~\cite{geraci_searching_2019,arvanitaki_sound_2016,Grote2019}. 
%for light with a frequency that is approximately independent of the changes in the fundamental constants. 

Laser interferometers for gravitational-wave (GW) detection are modified Michelson interferometers with exquisite sensitivity to differential changes in the optical path length of their arms. The thin cylindrical beamsplitter in such an instrument interacts asymmetrically with light from the two arms, as the front surface has a 50\% reflectivity and the back surface has an anti-reflective coating. Therefore, a change in the size ($\delta l$) and index of refraction ($\delta n$) of the beamsplitter affects the two arms differently, and produces an effective difference in the optical path lengths of the arms $L_{x,y}$
\begin{equation}\label{eq:deltaL}
    \delta(L_x - L_y)  \approx \sqrt{2}\left[\left(n-\frac{1}{2}\right)\delta l + l\delta n \right],
\end{equation}
\footnote{This expression includes a correction to Eq.~17 in \cite{Grote2019}. In addition, a geometrical correction ($\approx6.4\%$) from Snell's law is applied to Eqs.~\ref{eq:deltaL} and \ref{eq:full_signal} for calculating the results below.}
% in the adiabatic limit, which holds for a beamsplitter with a fundamental longitudinal vibrational frequency much larger than $\omega_\phi$ (the driving frequency)~\cite{Grote2019}. 
The mirrors in the arms of GW interferometers would also undergo changes in their size and index of refraction, but as the wavelength of the DM field is much greater than the distance between the arm mirrors ($\lambda_\phi/L\gtrsim 10^{3}$) for all frequencies of interest here, and because the mirrors have roughly the same thickness, the effect is almost equal in both arms and thus does not produce a dominant signal. 

The interferometer most sensitive to potential DM signals is the GEO\,600 detector, as it has the highest sensitivity to optical phase differences between the two arms. The squeezed vacuum states of light currently employed in this instrument allow for a world-record quantum noise reduction of 6~dB~\cite{GEO6dB_2021}. Although other GW  detectors (LIGO/Virgo) are more sensitive to gravitational waves through the use of Fabry-P\'erot cavities in the arms, these do not boost their sensitivity to signals induced at the beamsplitter, so their relative sensitivity to scalar DM is lower~\cite{Grote2019}. 

%The Fabry-P\'erot cavities used in the arms of most GW interferometers increase their sensitivity to gravitational waves, but do not increase the sensitivity to DM signals induced at the beamsplitter. Therefore, the interferometer most sensitive to scalar dark matter is the GEO\,600 detector, which has the highest sensitivity to optical phase differences between the two arms, and does not employ arm cavities \cite{Grote2019}. 
From Eqs.~\ref{eq:osc_field}--\ref{eq:deltaL} it follows that an oscillating scalar dark matter field is expected to produce a Doppler-shifted and -broadened signal in the GEO\,600 interferometer of the form
\begin{equation}\label{eq:full_signal}
    \delta (L_x - L_y) \approx \left(\frac{1}{\Lambda_\gamma} + \frac{1}{\Lambda_e} \right)\left(\frac{n\,l\,\hbar\,\sqrt{2\,\rho_{\mathrm{local}}}}{m_\phi\, c}\right)\cos\left(\omega_{\text{obs}} t\right),
\end{equation}
%where the contribution of changes of the beamsplitter's refractive index ($\delta n$) was neglected, as it is $\sim10^3$~times smaller than the contribution of the size changes of the beamsplitter ($\delta l$).
where we have neglected the contribution of the refractive index changes to the signal, as it is three orders of magnitude smaller than that of the size changes. Given this prediction, we can examine the data from the interferometer for the presence of such signals, and if none are found, place upper limits on the mass and coupling constants of scalar field DM.

% of  the GEO\,600 detector and  and in the absence  potential scalar DM signals, and  examined ruling out the presence of such a signal in data from the GEO\,600 detector therefore allows us to set upper limits on the mass and coupling constants of scalar field DM.% set constraints on the properties of scalar dark matter.  

% for $\omega > \omega_{0,e}$, where $\omega$ is the angular frequency of the light and $\omega_{0,e}$ is the electronic resonance of the solid and   using a light source with an angular frequency $\omega$ that is independent of the changes in the fundamental constants, and 

% \input{sections/sec_Methods}
% \input{sections/sec_Results}
\section{Results}\label{sec:results}
% \begin{figure*}[ht]
% \centering
% \includegraphics[scale=0.34]{figures/AmplSpec_PaperR1_landscp_Lambda_Excl._06-03-2021_1459.pdf}
% \caption{A typical amplitude spectrum (black) produced with frequency bins that are tuned to the expected dark matter linewidth using the modified LPSD technique. The noise spectrum was estimated at each frequency bin from neighbouring bins to yield the local noise median (blue) and 95\% confidence level (green). Peaks (red) above this confidence level were considered candidates for DM signals and subjected to follow-up analysis.}
% \label{fig:peak_selection}
% \end{figure*}

\begin{figure}[ht]
% Non-Nature Format:
%\includegraphics[width=0.5\textwidth]{figures/AmplSpec_PaperR1_square_Lambda_Excl._06-08-2021_1804.pdf}
% Nature Format:
\includegraphics[width=0.49\textwidth]{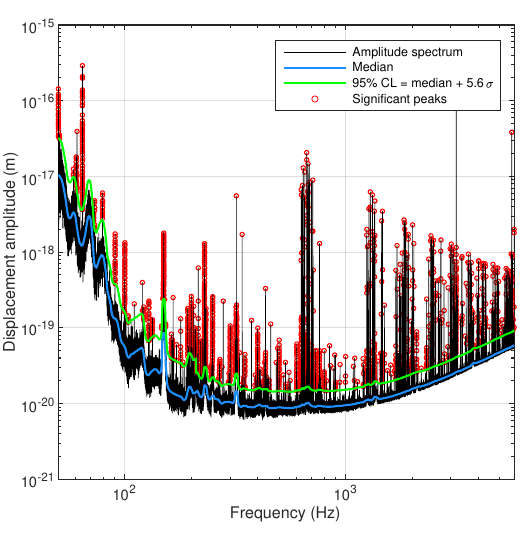}
\caption{A typical amplitude spectrum (black) produced with frequency bins that are tuned to the expected dark matter linewidth using the modified LPSD technique. The noise spectrum was estimated at each frequency bin from neighbouring bins to yield the local noise median (blue) and 95\% confidence level (green). Peaks (red) above this confidence level were considered candidates for DM signals and subjected to follow-up analysis.}
\label{fig:peak_selection}
\end{figure}
\begin{figure}[ht]
% Non-Nature Format:
%\includegraphics[width=0.49\textwidth]{figures/BasicScalar_PaperR1.1_square_Lambda_Excl._06-10-2021_1542.pdf}
% Nature Format:
%\includegraphics[width=0.49\textwidth]{figures/BasicScalar_PaperR2.0_square_Lambda_Excl._08-26-2021_1108.pdf}
\includegraphics[width=0.49\textwidth]{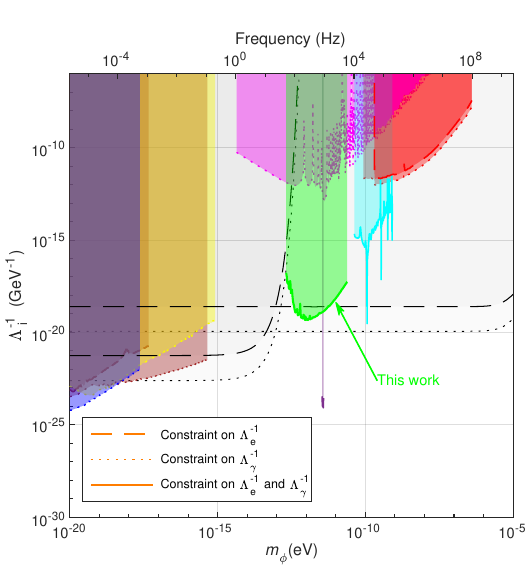}
\caption{Constraints on the coupling parameters $\Lambda_\gamma$, $\Lambda_e$ as a function of the field's mass $m_\phi$, for scalar field DM as in the \textit{Basic Scalar} scenario (see text). Dashed lines represent constraints on the electron coupling $\Lambda_e$ and dotted lines represent constraints on the photon coupling $\Lambda_\gamma$, at the 95\% confidence level. The green region denotes the parameter space excluded in the current study through the spectral analysis of data from the GEO\,600 gravitational-wave detector. Other coloured regions indicate parameter space excluded through previous direct experimental searches; to wit, Hees et al~\cite{hees_searching_2016}~(blue), Van Tilburg~et~al~\cite{van_tilburg_search_2015}~(yellow), Kennedy~et~al~\cite{kennedy_precision_2020}~(brown), Aharony~et~al~\cite{aharony_constraining_2019}~(magenta), Branca~et~al~\cite{branca_search_2017}~(purple), Savalle~et~al~\cite{savalle_searching_2021}~(cyan), and Antypas~et~al~\cite{antypas_scalar_2019}~(red)). The black curves and grey regions correspond to previous constraints from `fifth-force' (FF) searches/tests of the equivalence principle (EP); to wit, the most stringent such constraints for this DM scenario are from the MICROSCOPE experiment~\cite{berge_microscope_2018,leefer_search_2016} (lower curves at low mass), and the Cu/Pb torsion pendulum experiment performed by the E\"{o}t-Wash group~\cite{smith_short-range_1999,wagner_torsion-balance_2012,hees_violation_2018} (at higher masses).}
\label{fig:constraints_BS}
\end{figure}

%\subsection{Spectral estimation}
The GEO\,600 interferometer~\cite{Dooley_2016} has been in joint observing runs with the Advanced LIGO detectors since 2015, primarily to look for gravitational waves.
%The GEO\,600 interferometer~\cite{Dooley_2016} has been in observing runs since 2015 together with the Advanced LIGO detectors to primarily look for gravitational waves.
We performed spectral analysis on seven $T\sim 10^5$~s segments of strain data from the GEO\,600 interferometer (acquired in 2016 and 2019) using a modified version of the LPSD technique ~\cite{trobs_improved_2006,*trobs2009improved}, which was designed to produce spectra with logarithmically spaced frequencies. Using this algorithm to perform discrete Fourier transforms (DFTs) with a frequency dependent length, we created spectra in which each frequency bin was made to have a width equal to the Doppler-broadened linewidth of potential signals from scalar field DM in a galactic halo. This method yields in theory the maximum attainable signal-to-noise ratio (SNR), given a certain amount of data (see Sec.~\ref{sec:nature_meth})~\cite{miller_adapting_2020,derevianko_detecting_2018}. A matched filtering approach is not feasible as the phase of the signal varies stochastically.

%\subsection{Candidate signal search} \label{method-signalsearch}
We analysed the amplitude spectra of all seven strain data segments %of length $T\sim 10^5$~s 
for the presence of DM signals by looking for significant peaks in the underlying noise. Peaks were considered candidates when there was a less than $5\%$ probability of the local maximum being due to noise, where we compensated for the look-elsewhere effect using a large trial factor ($\sim10^{6}$). 
% To determine this probability for every peak, the noise was assumed to be Gaussian with a frequency-dependent expectation value and variance. The local noise parameters were estimated at every frequency bin from $w=5\cdot10^4$ neighbouring bins. This method allows the underlying noise distribution to be estimated in a way that is independent of narrow ($\ll w$) spectral features (such as those due to mechanical excitation of the mirror suspensions), under the assumption that the underlying noise spectrum is locally flat (that is, the auto-correlation length of the noise spectrum is assumed to be $\gg w$)\cite{Note_on_w}.

This analysis found %$\approx 3.5\pm 0.5\cdot 10^4$ 
$\sim 10^4$ peaks above the 95\% confidence level ($\gtrsim 5.6 \sigma$), where the total error includes a frequency dependent amplitude calibration error of up to $30\%$ inherent to GEO\,600 data \cite{GEOcal_2019}. The frequency and amplitude stability of the peaks in time was then evaluated by cross-checking all candidates between spectra. Candidate peaks were rejected if their centre frequencies differed between spectra by more than the Doppler shift expected from the Earth's motion around the Sun through a galactic DM halo \cite{freese_annual_2013}. Peaks were also rejected if their amplitude changed significantly ($\gtrsim 5 \sigma$) between spectra. 
%between spectra by more than an amount expected due to the underlying noise. Specifically, peaks that changed amplitude between spectra were only eliminated when there was a less than $1\%$ ($\gtrsim 5 \sigma$) probability that the amplitude change was due to noise (compensated for the look-elsewhere effect).%, so that on average only one peak would be falsely rejected if all cross-checks are performed 100 times). 

Using this procedure, we eliminated all but 14 candidate peaks, where the vast majority ($>99\%$) of peaks were rejected because they did not appear in all data sets within the centre frequency tolerance.

%\subsection{Follow-up analysis of candidates}
These 14 candidate peaks were subjected to further analysis to investigate if their properties matched that of a DM signal. 13 of the peaks were found to have insufficient width to be caused by DM ($\Delta f_\mathrm{peak}/\Delta f_\mathrm{DM}\lesssim 10$, see Sec.~\ref{sec:nature_meth}). %\vivien{This is a good argument, but it has a significant impact on the results. So I think it'd be best to give number, what's the width of those peaks (average, maximum) compared to the smallest width one can expect from DM, to help the reader get behind this argument}.
% Investigation of each of these candidates found that shifting the bin centre frequencies by an amount much smaller than the expected linewidth of DM signals of that frequency and amplitude and recomputing the spectra did not reproduce the peak \cite{Note_on_peaks}. Specifically, the ratio of observed and expected peak width was $\Delta f_\mathrm{peak}/\Delta f_\mathrm{DM}\lesssim 10$ for these 13 peaks. %\footnote{Additional work revealed these 20 candidate peaks were not present in spectra created using the same data and the same LPSD algorithm implemented in a different programming language, whereas the noise floor and other spectral features were reproduced identically. These peaks are therefore likely artefacts of the numerical implementation of the LPSD technique.}.
% The remaining candidate peak was also ultimately rejected, as although it appeared to have sufficient frequency spread to be a DM signal, additional analysis showed this signal has a coherence time much greater than that expected for a DM signal of that frequency ($\tau_c^\mathrm{peak}/\tau_c^\mathrm{DM}>10$, see Sec.~\ref{sec:nature_meth}).
The remaining candidate peak had sufficient frequency spread to be due to DM, but additional analysis showed this signal has a coherence time much greater than that expected for a galactic halo DM signal of that frequency ($\tau_c^\mathrm{peak}/\tau_c^\mathrm{GH}>10$, see Sec.~\ref{sec:nature_meth}). This leaves open the possibility of the signal being due to scalar DM gravitationally bound to Earth, such as in a Relaxion halo. However, additional investigations revealed this signal was not present in data acquired with auxiliary electronics, whereas the noise and the other signals from the interferometer were. 
%This fact, in combination with high-resolution spectra revealing that the frequency at which the peak occurs is very close to and indistinguishable from an integer ($f_\mathrm{peak} = 224 \pm 2\cdot 10^{-5}$~Hz), imply the signal is most likely an artefact of a timing signal in the main data acquisition electronics. 
% ; i.e. the height of the peak did not decrease for DFT lengths more than an order of magnitude greater than the expected DM coherence time.
The peak was therefore rejected, and is suspected to be an artefact from a timing signal in the main data acquisition electronics (see \ref{sec:nature_meth} for details).

\begin{figure*}[!htb]
  \begin{minipage}[b]{0.49\textwidth}
    % \includegraphics[width=\textwidth]{figures/Dilaton_PaperR1.1_square_Lambda_Excl._06-10-2021_1539.pdf}
    %Nature Format:
    \includegraphics[width=\textwidth]{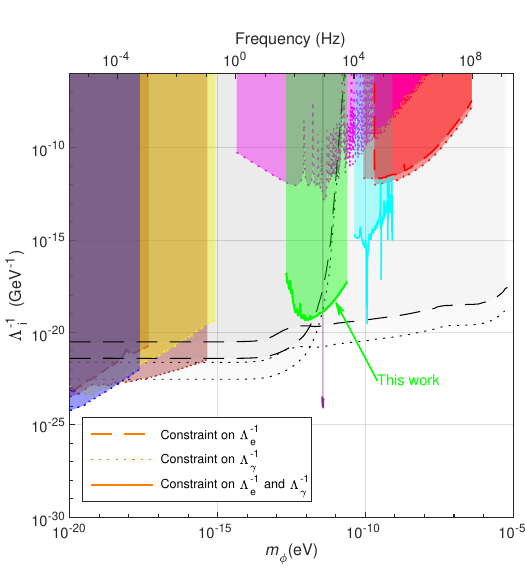}
    %\caption{}
  \end{minipage}
  \hfill
  \begin{minipage}[b]{0.49\textwidth}
    % \includegraphics[width=\textwidth]{figures/RelaxionH_PaperR1.1_square_Lambda_Excl._06-10-2021_1538.pdf}
    %Nature Format:
    \includegraphics[width=\textwidth]{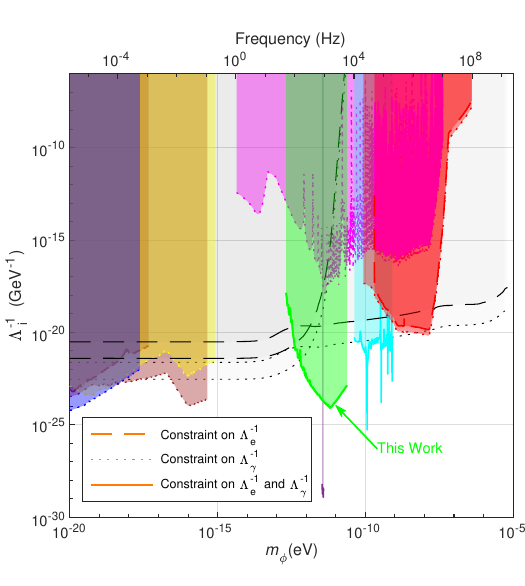}
    %\caption{}
  \end{minipage}
  \caption{Constraints on the coupling parameters $\Lambda_\gamma$, $\Lambda_e$ as a function of the field's mass $m_\phi$, for scalar field DM as in the \textit{Dilaton/Modulus} scenario (\textbf{left}) and the \textit{Relaxion Halo} scenario (\textbf{right}). Dashed lines represent constraints on the electron coupling $\Lambda_e$ and dotted lines represent constraints on the photon coupling $\Lambda_\gamma$, at the 95\% confidence level. The green region denotes the parameter space excluded in the current study through the spectral analysis of data from the GEO\,600 gravitational-wave detector. Other coloured regions indicate parameter space excluded through previous direct experimental searches~(\cite{van_tilburg_search_2015,hees_searching_2016,kennedy_precision_2020,aharony_constraining_2019,savalle_novel_2019,antypas_scalar_2019, branca_search_2017},~see caption of Fig.~\ref{fig:constraints_BS}). The black lines and grey regions correspond to previous constraints from `fifth-force' (FF) searches/tests of the equivalence principle (EP); to wit, the most stringent such constraints for this DM scenario are from the MICROSCOPE experiment~\cite{berge_microscope_2018,hees_violation_2018} (lower curves at low mass), and the Be/Ti torsion pendulum experiment performed by the E\"{o}t-Wash group~\cite{schlamminger_test_2008,hees_violation_2018} (at higher masses). The constraints for the \textit{Relaxion Halo} scenario from direct experimental searches have been obtained by rescaling the originally reported constraints to account for the mass-dependent local overdensities as proposed in \cite{banerjee_relaxion_2020}. This produces novel constraints not reported before for \textit{Relaxion Halo} DM from the results of \cite{van_tilburg_search_2015,hees_searching_2016,aharony_constraining_2019,kennedy_precision_2020,branca_search_2017}. The FF/EP constraints are independent of the local DM density and are thus unchanged.}
  \label{fig:constraints_DM&RH}
\end{figure*}

Having determined that all significant peaks in the amplitude spectrum %that are above the noise at a confidence level of $>99\%$
are not caused by scalar field DM, we can set constraints on the parameters of such dark matter at a 95\% confidence level (corresponding to $5.6\sigma$ above the noise floor), using Eq.~\ref{eq:full_signal}. We apply our results to three different scalar DM scenarios considered in literature: 
\begin{enumerate}
    \item \textit{Basic Scalar} (Fig.~\ref{fig:constraints_BS}): The scalar field DM is assumed to interact with the SM as given by the terms in~\ref{L_int}, and is further assumed to be homogeneously distributed over the solar system with a density of $\rho_{\mathrm{GH}}=0.4~\mathrm{GeV}/\mathrm{cm}^3$, as in the standard galactic DM halo model \cite{read_local_2014}.
    \item \textit{Dilaton/Modulus} (Fig.~\ref{fig:constraints_DM&RH}, left): In addition to the coupling to the electromagnetic sector as in Eq.~\ref{L_int}, the field is assumed to have couplings to the QCD sector, and the coupling to the gluon field is assumed to be dominant \cite{arvanitaki_sound_2016,damour_string_1994,damour_equivalence_2010,wagner_torsion-balance_2012}. The local DM density is taken to be $\rho_\mathrm{GH}$. Compared to the \textit{Basic Scalar}, this scenario is subject to additional limits from tests of the equivalence principle, but is equally constrained by our result and those of other direct searches.
    \item \textit{Relaxion Halo} (Fig.~\ref{fig:constraints_DM&RH}, right): In this scenario, the scalar field effectively couples to the SM as in the \textit{Dilaton/Modulus} scenario, but these couplings arise through mixing with the Higgs boson \cite{flacke_phenomenology_2017,banerjee_coherent_2019}. The local DM density in this scenario is taken to be dominated by a Relaxion halo gravitationally bound to earth, which leads to a local overdensity that depends on the field's mass and reaches values of up to $\rho_\mathrm{local}/\rho_\mathrm{GH} = 10^{11}$ for the mass range constrained in this work \cite{banerjee_relaxion_2020}. 
\end{enumerate}

For each scenario, we set constraints on the electron and photon coupling parameters $\Lambda_e$, $\Lambda_\gamma$, as a function of the field's mass $m_\phi$ (where for each coupling constant we assume the other to be zero); the constraints are plotted in Figs.~\ref{fig:constraints_BS} and \ref{fig:constraints_DM&RH} together with previous upper limits. For the \textit{Relaxion Halo} scenario, we assumed a mass-dependent halo density as described in \cite{banerjee_relaxion_2020}. 

Constraints from other direct experimental DM searches include those from various atomic spectroscopy experiments~\cite{van_tilburg_search_2015,hees_searching_2016,aharony_constraining_2019,aharony_constraining_2019,antypas_scalar_2019}, a search using an optical cavity~\cite{savalle_searching_2021}, and a resonant mass detector~\cite{branca_search_2017}. Tests of the equivalence principle (EP) using e.g. torsion balances \cite{smith_short-range_1999,schlamminger_test_2008,wagner_torsion-balance_2012} have also been used to set constraints on the parameters of undiscovered scalar fields; these bounds assume the scalar field manifests as a `fifth force' (FF), and is sourced by a test mass (e.g. the Earth) \cite{berge_microscope_2018,hees_violation_2018,leefer_search_2016}. The constraints on scalar fields inferred from these experiments depend in general on the composition and topography of the test masses and are independent of the local dark matter density.  
% are inferred from data from various torsion balance experiments (done by the E\"{o}t-Wash group)  and the space-based MICROSCOPE experiment \cite{berge_microscope_2018},
% These constraints cover the mass range between $\approx {2\cdot10^{-13}\;\mbox{--}\;2\cdot10^{-11}}$~eV, and improve over the current upper limits on both $\Lambda_e$ and $\Lambda_\gamma$ set through atomic spectroscopy experiments \cite{aharony_constraining_2019,stadnik_can_2015,stadnik_enhanced_2016} by more than six orders of magnitude. In addition,  our results constrain the scalar DM-electron coupling constant $\Lambda_e$ at a level one order of magnitude lower than previous bounds from tests of the equivalence principle \cite{smith_short-range_1999,hees_violation_2018}. 
\section{Conclusions}\label{sec:conclusions}
In this paper, we presented the first search for signals of scalar field dark matter in the data of a gravitational-wave detector. Scalar field dark matter would cause oscillations of the fine structure constant and electron mass, which in turn drive oscillations of the size and index of refraction of the beamsplitter in an interferometer. This would thus produce an oscillatory signal in a gravitational-wave detector at a frequency set by the mass of the dark matter particle.

As exquisite classical noise mitigation is employed in gravitational-wave detectors, quantum technologies such as squeezed light can provide a major increase in sensitivity. Such technologies facilitate measurements beyond the shot-noise quantum limit, and yield unprecedented sensitivity to scalar field dark matter in a wide mass range.
% The exquisite noise mitigation employed in gravitational-wave detectors allows for the application of quantum technologies such as squeezed light, facilitating measurements beyond the shot-noise quantum limit, at unprecedented sensitivity to scalar field dark matter in a wide mass range.

In addition, by tuning the frequency bin widths to the expected dark matter linewidth, our spectral analysis method improves on the analyses used in previous work that set constraints on dark photons using data from gravitational-wave detectors, and other searches for scalar fields in frequency space. In contrast to these other efforts, the spectral analysis presented here yields the optimal signal-to-noise ratio for potential dark matter signals across the full frequency range. 

We excluded the presence of such signals in the data of the GEO\,600 gravitational-wave detector, thereby setting new lower limits on dark matter couplings at up to $\Lambda_{e},\Lambda_{\gamma} = 3\cdot10^{19}$~GeV for dark matter masses between $10^{-13}$ and $10^{-11}$~eV. The new constraints improve upon the current limits in this mass range obtained with atomic spectroscopy experiments by more than six orders of magnitude, and are up to four orders of magnitude more stringent than previous bounds from tests of the equivalence principle for some dark matter scenarios. 
%  These constraints cover the mass range between $\approx {2\cdot10^{-13}\;\mbox{--}\;2\cdot10^{-11}}$~eV, and improve over the current upper limits on both $\Lambda_e$ and $\Lambda_\gamma$ set through atomic spectroscopy experiments \cite{aharony_constraining_2019,stadnik_can_2015,stadnik_enhanced_2016} by more than six orders of magnitude. In addition,  our results constrain the scalar DM-electron coupling constant $\Lambda_e$ at a level one order of magnitude lower than previous bounds from tests of the equivalence principle \cite{smith_short-range_1999,hees_violation_2018}.

Tighter constraints on scalar field dark matter in various mass ranges can be set in the future using new yet-to-be-built gravitational-wave detectors or other similar precision interferometers. Using the same methods as in this work these instruments would allow new limits to be set across their characteristic sensitive frequency range. Moreover, by slightly modifying the optics in such interferometers, e.g. by using mirrors of different thicknesses in each interferometer arm, their sensitivity to scalar field dark matter could be improved even further~\cite{Grote2019}.
%By lowering losses, quantum technologies such as squeezed light are expected to progress as well, becoming an indispensable tool for fundamental physics research. 
% Through the reduction of losses, quantum technologies such as squeezed light are also expected to improve, making them an indispensable tool for fundamental physics research. 
Through the reduction of losses, quantum technologies such as squeezed light are also expected to improve, allowing for ever-increasing noise mitigation \cite{pradyumna_twin_2020}. These and other forthcoming technological advances make precision interferometers operating beyond quantum limits indispensable tools %very promising experimental means 
for dark matter detection and fundamental physics in general.

%More stringent limits can theoretically be set by analysing more data. This allows for more averaging thus decreasing the variance of the spectrum proportional to the inverse of the square root of the amount of data, such that the constraint converges asymptotically to the mean noise level. Constraints can be set beyond this level using longer DFT lengths at the cost of reduced SNR, but this is subject to severely diminishing returns; the constraint can only be improved by a factor proportional to the fourth root of the amount of data needed \cite{derevianko_detecting_2018} (and the computation time scales with the product of DFT length and the amount of data ~\cite{trobs_improved_2006, trobs2009improved}).

%\input{sections/sec_Method_Validation}
%\input{sections/alternatives}
%\input{sections/scratch}
%\input{sections/section02.tex}
%\input{sections/section03.tex}
\section*{Acknowledgements} \label{sec:acknowledgements}
The authors thank Yevgeny Stadnik and Yuta Michimura for valuable discussion and comments on this work. 
We thank Duncan Macleod and Paul Hopkins for significant programming assistance, and
Michael Tr\"obs and Gerhard Heinzel for permission to use their LPSD code.
The authors are grateful for support from
the Science and Technology Facilities Council (STFC), grants
ST/T006331/1, ST/I006285/1, and ST/L000946/1, 
the Leverhulme Trust, grant RPG-2019-022, and
the Universities of Cardiff and Glasgow in the United
Kingdom, 
the Bundesministerium für Bildung und
Forschung, the state of Lower Saxony in Germany, the
Max Planck Society, Leibniz Universit\"at Hannover, and
Deutsche Forschungsgemeinschaft (DFG, German
Research  Foundation) under Germany’s  Excellence
Strategy EXC 2123 QuantumFrontiers 390837967.
This work also was partly supported by DFG grant SFB/
Transregio 7 Gravitational Wave Astronomy.
We further thank Walter Grass for his
years of expert infrastructure support for GEO\,600.
This document has been assigned LIGO document number LIGO-P2100053.
% only for nature submission:
\section*{Author contributions}
S.M.V. and P.R. have analysed the data and compiled the results; H.G. has instigated this work and S.M.V. and H.G. have written
the manuscript; V.R has given critical input to the analysis. 
C.A., J.L. and K.D. have lead the GEO\,600 instrument group during the period where data for this work was acquired. 
F.B., A.B, S.D. H.L., N.M., S.N., E.S, B.S., K.A.S, M.B., V.K, M.W., and H.W.
have worked on the instrument in different capacities required to achieve sensitivity and extended run duration; 
B.W. has provided laser expertise and H.V. and M.M. have built the squeezed-light source.
\section{Methods}\label{sec:nature_meth}
\small

\begin{figure*}[!htb]
  \begin{minipage}[b]{0.43\textwidth}
    \includegraphics[width=\textwidth]{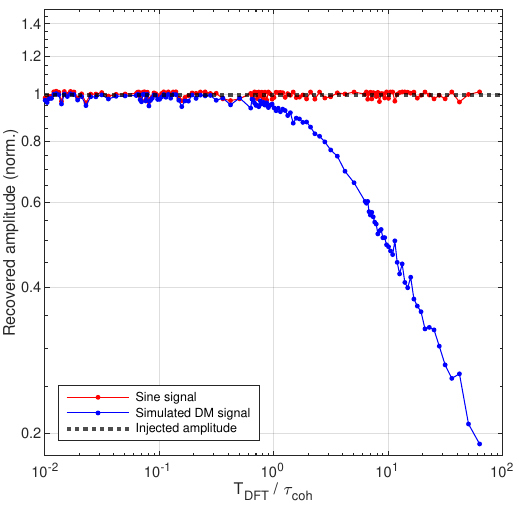}
    %\caption{}\label{fig:Ampl_scaling}
  \end{minipage}
  \hspace{10mm}
  \begin{minipage}[b]{0.43\textwidth}
    \includegraphics[width=\textwidth]{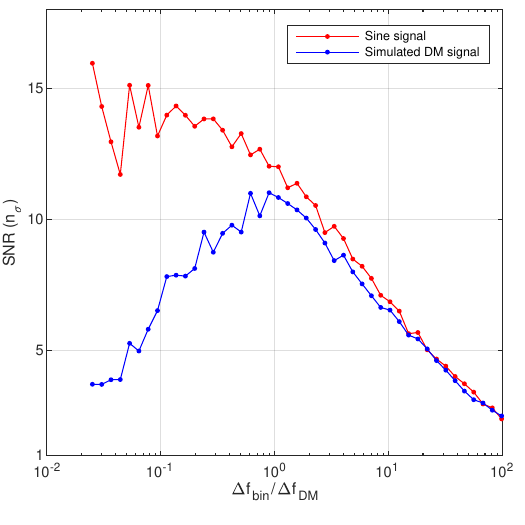}
    %\caption{}\label{fig:SNR_scaling}
  \end{minipage}
  \caption{The spectral amplitude (left) and signal-to-noise ratio (SNR, right) of a simulated DM signal (blue) and monochromatic sine wave (red) as recovered from spectra created using different frequency bin widths ($\Delta f_\mathrm{bin}=1/T_\mathrm{DFT}$). The plotted recovered amplitude is normalised by the injected amplitude. The SNR ($n_\sigma$) is measured as the difference between the signal amplitude and the noise amplitude divided by the standard deviation of the noise. The appearance of a maximum for the SNR as shown on the right is a direct consequence of both the decrease of the recovered amplitude of signals with limited coherence (as shown on the left) and the scaling of white Gaussian noise with increasing integration time. The plot on the left was produced by injecting a simulated dark matter signal and a perfect sine into a segment of GEO\,600 data and creating spectra using the modified LPSD technique described above. The plot on the right was made by injecting the same signals into white Gaussian noise and creating spectra using Welch's method. Note that for any single bin and for equal $T_\mathrm{DFT}$ the spectral estimate obtained with the LPSD method (Eq.~\ref{eq:LPSD_spec_est}) is mathematically equal to that obtained with Welch's method.}
  \label{fig:amplSNR_scaling}
\end{figure*}

\subsection{Spectral estimation}
Spectral analysis was performed using a modified version of the LPSD technique ~\cite{trobs_improved_2006,*trobs2009improved}. This technique is designed to produce spectral estimates with logarithmically spaced frequencies, and thus allows for the production of spectral estimates with a frequency-dependent bin width. Using this technique, we subdivided the $\sim10^5$~s data segments into
\begin{equation}
    N_f=\left\lfloor \frac{T-\tau_\text{coh}(f)}{\tau_\text{coh}(f) (1-\xi)} +1 \right\rfloor  
\end{equation}
smaller overlapping subsegments $S_f^k(t)$ with a length equal to the expected coherence time $\tau_\text{coh}(f)$, of a dark matter (DM) signal at a frequency $f$, where $\xi\in[0,1]$ is the fractional overlap of the subsegments, and $k\in [1, N_f]$). As the expected coherence time and linewidth is frequency dependent, this subdivision is unique for every frequency of interest. After subdivision, the subsegments were multiplied with a Kaiser window function $W_f(t)$ and subjected to a DFT at a single frequency:
\begin{equation}\label{eq:LPSD_spec_est}
    a^k(f) = \sum_{t=0}^{T_\mathrm{DFT}} W_f(t)\, S_f^k(t)\, \mathrm{e}^{2\pi i f t}, 
\end{equation}
with $T_\mathrm{DFT}=\tau_\text{coh}(f)$, where $a^k(f)$ is thus the complex spectral estimate at frequency $f$ for the $k^\text{th}$ subsegment. Frequency points are chosen by dividing the interval between the chosen minimum frequency (50~Hz) and the Nyquist frequency ($\approx8.2$~kHz) by the DM linewidth, and then rounding the resulting number of bins to the nearest integer to set the final frequency points and bin widths. The absolute squared magnitudes $|a^k(f)|^2$ are averaged over the subsegments to obtain the power spectrum 
\begin{equation}
P(f)=\frac{C}{N_f}\sum_{k=1}^{N_f}|a^k(f)|^2,
\end{equation}
where $C$ is a normalisation factor. The spectra used in the analysis were made with a bin width equal to the expected linewidth of galactic DM ($\Delta \omega /\omega\approx10^{-6}$, see \cite{derevianko_detecting_2018}).  The amplitude spectrum $A(f)= \sqrt{P(f)}$ created in this way comprises $\approx5\cdot10^6$ frequency bins between 50 Hz and 6 kHz.

%The SNR for DM signals in such a spectrum is optimal given a certain amount of data (see \ref{sec:methodval}), and can only be further improved by analysing more data, which allows for more averaging thus decreasing the variance of the spectrum proportional to the inverse of the square root of the amount of data, such that the sensitivity approaches the noise floor.
The SNR for galactic DM signals in such a spectrum is optimal given a certain amount of data (see Sec.~\ref{sec:methodval}), and can only be further improved by analysing more data. Additional data would allow for more averaging, which decreases the variance of the spectrum as the square root of the amount of data, such that the sensitivity approaches the noise floor.
The noise floor can be lowered using longer DFT lengths at the cost of reduced SNR, but this is subject to severely diminishing returns; the sensitivity can only be improved by a factor proportional to the fourth root of the amount of data needed \cite{derevianko_detecting_2018} (and the computation time scales with the product of DFT length and the amount of data \cite{trobs_improved_2006}).  Computation times for the spectra used in this work are ${\sim 10}$~s per frequency bin for each ${\sim 10^5}$~s data set, or ${\sim10^4}$~CPU hours per spectrum.

\subsection{Estimation of noise statistics} 
The local noise parameters were estimated at every frequency bin from $w=5\cdot10^4$ neighbouring bins. This method allows the underlying noise distribution to be estimated in a way that is independent of narrow ($\ll w$) spectral features (such as those due to mechanical excitation of the mirror suspensions), under the assumption that the underlying noise spectrum is locally flat (that is, the auto-correlation length of the noise spectrum is assumed to be $\gg w$). The choice of $w$ thus represents a trade-off between erroneously assuming instrumental spectral artefacts or signals to be features of the underlying noise spectrum versus erroneously assuming features of the underlying noise spectrum to be instrumental spectral artefacts or signals.

\subsection{Follow-up analysis of candidates}
As mentioned above, 14 candidate peaks remained after cross-checking spectra taken at different times. 13 of these peaks were found to have insufficient width to be DM signals. Further investigation of each of these candidates found that shifting the bin centre frequencies by an amount much smaller than the expected linewidth of DM signals of that frequency and amplitude and recomputing the spectra did not reproduce the peak. Additional work revealed these 13 candidate peaks were not present in spectra created using the same data and the same LPSD algorithm implemented in a different programming language, whereas the noise floor and other spectral features were reproduced identically. These peaks are therefore likely artefacts of the numerical implementation of the LPSD technique. 

The coherence time of the single remaining candidate peak was probed by evaluating its height in the amplitude spectrum as a function of the DFT length (see Sec.~\ref{sec:methodval}). The height of the peak did not decrease for DFT lengths more than an order of magnitude greater than the expected DM coherence time, evidencing a coherence time much greater than that expected for a galactic DM signal of that frequency. To find the origin of the signal, and to check whether it could be due to the theoretically more coherent \textit{Relaxion Halo} DM, we performed spectral analysis on data acquired on an auxiliary data acquisition system. The signal was not present in this data, whereas both noise and other signals from the interferometer were. This fact, in combination with high-resolution ($\Delta f/f \sim 10^{-7}$) spectra revealing that the frequency at which the peak occurs is very close to and indistinguishable from an integer ($f_\mathrm{peak} = 224 \pm (2\cdot 10^{-5})$~Hz), implies the signal is most likely an artefact of a timing signal in the main data acquisition electronics.

\subsection{Validation of methods}\label{sec:methodval}
To validate several aspects of our analysis methods, we simulated DM signals and injected them into sets of real and simulated data. The DM signals were created by superposing $\sim10^2$ sinusoids at frequencies linearly spaced around a centre frequency (the simulated Doppler-shifted DM Compton frequency), where the amplitude of each sinusoid is given by the quasi-Maxwellian DM line shape proposed in \cite{derevianko_detecting_2018} scaled by a simulated DM coupling constant; the relative phases of the sinusoids are randomised to capture the thermalisation of the scalar field DM.

To test the spectral estimation, signal search, and candidate rejection, a blind injection of simulated DM signals into several GEO\,600 data sets was performed, where the frequency, amplitude, and number of signals was masked to the authors. All injected signals were recovered at their Compton frequency and at an amplitude corresponding to the hypothetical coupling constant, and were subsequently identified through cross-checks between spectra as persistent candidate DM signals.

The formerly proposed \cite{derevianko_detecting_2018,pierce_searching_2018} and herein utilised condition of setting the frequency bin widths equal to the expected DM line width for attaining optimal SNR was tested using simulated DM signals as well. Mock DM signals and monochromatic sine signals were injected into real GEO\,600 data and Gaussian noise, and spectra were made for which the width of the frequency bins $\Delta f_\mathrm{bin}$ (and correspondingly the length of the DFTs $T_\mathrm{DFT}$)  was varied over four orders of magnitude. The recovered amplitude of signals injected into GEO\,600 data in spectra created using the LPSD algorithm is plotted in Fig.~\ref{fig:amplSNR_scaling} (left). This shows that the recovered amplitude of signals starts to decrease as the DFT length exceeds the coherence time (a monochromatic sine has infinite coherence time), and validates the rejection of the remaining candidate signal above as its amplitude was found to be roughly constant for $T_\mathrm{DFT}/\tau_c>10$. The recovered SNR of signals injected into Gaussian noise in spectra created using Welch's method \cite{Welch_1967} is plotted in Fig.~\ref{fig:amplSNR_scaling} (right), which confirms that the SNR is maximal when the frequency bin width is roughly equal to the full-width at half-maximum $\Delta f_\mathrm{DM}$ of the spectral line shape of the signal. This is a consequence of the aforementioned decrease in recovered amplitude for smaller bin widths and the scaling of white Gaussian noise.

\section{Data Availability} \small
%The data central to the results of this manuscript are available from the corresponding author upon reasonable request.
%The raw time series data used for the analysis are currently only available to members of the LIGO Scientific Collaboration, but will be made publicly available soon, in accordance with the LIGO Data Management Plan (\url{https://dcc.ligo.org/public/0009/M1000066/025/LIGO-M1000066-v25.pdf}). The upper limits in Fig.~\ref{fig:constraints} and intermediate results are available from the corresponding author upon reasonable request.
The upper limit data in Figs.~\ref{fig:constraints_BS} and \ref{fig:constraints_DM&RH} and intermediate results, such as the spectrum in Fig.~\ref{fig:peak_selection}, are available from the corresponding author upon request. The raw data used for the full analysis comprises about 80\,GB and is available from the corresponding author upon reasonable request.

\section{Code Availability} \small
The code used for this analysis has been released and can be found at the following location \url{https://github.com/philrelton/Scalar-Dark-Matter-LPSD}

%\begin{thebibliography}{4}
%\end{thebibliography}

\bibliographystyle{ieeetr} % puts references in order of citation sequence, but no name abbreviations
\bibliography{bibfile}

%\appendix*
%\input{sections/appendix1.tex}

\end{document}